# Spectroscopy of the Methane $\nu_3$ Band with an Accurate Mid-Infrared Coherent Dual-Comb Spectrometer

E. Baumann, F. R. Giorgetta, W. C. Swann, A. M. Zolot, I. Coddington, and N. R. Newbury

*National Institute of Standards and Technology, 325 Broadway, Boulder, Colorado 80305*

We demonstrate a high-accuracy dual-comb spectrometer centered at 3.4 µm. The amplitude and phase spectra of the P, Q, and partial R-branch of the methane $\nu_3$ band are measured at 25 MHz to 100 MHz point spacing with ~kHz resolution and a signal-to-noise ratio of up to 3500. A fit of the absorbance and phase spectra yield the center frequency of 132 rovibrational lines. The systematic uncertainty is estimated to be 300 kHz, which is $10^{-3}$ of the Doppler width and a tenfold improvement over Fourier transform spectroscopy. These data are the first high-accuracy molecular spectra obtained with a direct comb spectrometer.

PACS numbers 33.20.Ea, 42.62.Eh, 07.57.Ty

# I. INTRODUCTION

Frequency combs are attractive sources for spectroscopy because of their broadband, collimated output, and because the frequency of each discrete comb tooth can be referenced to an established frequency standard [1, 2]. Frequency combs can support conventional saturated absorption spectroscopy by providing a frequency ruler against which to measure the cw laser frequency [3-5]. They can also directly probe molecules over a broad spectrum [6-20]. Many demonstrations of direct comb spectroscopy have focused on speed, spectral bandwidth or sensitivity, with the frequency calibration based on a known spectral line rather than the underlying comb. Our focus here is instead on the high accuracy and resolution possible with a direct comb spectrometer. We require, at a minimum, that the spectrometer resolve individual teeth of a comb and that the comb is referenced to an accurate frequency standard.

Dual-comb spectroscopy is one method of direct comb spectroscopy that is particularly well-suited to this task. In a dual-comb spectrometer, a sensing comb is transmitted through a sample and then heterodyned against a local oscillator (LO) comb which has a repetition rate that differs by $\Delta f_r$ from the sensing comb repetition rate [9-20]. The basic concept is illustrated in Fig. 1 for the case of ideal frequency combs. In the frequency domain, the result is a rf comb where each rf tooth maps directly to a single tooth of the sensing comb. In the time-domain, the pulse train from the LO comb walks through the sensing pulse train to create an interferogram every $1/\Delta f_r$ in analogy with cross-correlation measurements or Fourier transform infrared spectrometers (FTIR); the Fourier transform of a long series of such interferograms leads again to the rf comb of Fig. 1.

Dual-comb spectroscopy has been implemented across the extremes ranging from free-running combs to phase-coherent, frequency-stabilized combs [9-20]. In the former, the measurement is essentially a single cross-correlation measurement between the two pulse trains; there is no coherence from one interferogram to the next so that the rf comb of Fig. 1 (and therefore the optical comb) is not resolved, nor is there intrinsic absolute frequency accuracy. It does, however, avoid the experimental complexity associated with phase-locked, referenced frequency combs. In the latter implementation, the combs are tightly phase-locked together for high mutual coherence and residual linewidths well below $\Delta f_r$. The combs are also stabilized to an absolute frequency reference. In this case, the rf comb (and therefore optical comb) is fully resolved and the stabilized comb accuracy can be applied to the frequency axis of the recorded spectra. Furthermore, the signal can be integrated over multiple interferograms for times much longer than $1/\Delta f_r$, limited only by the effective mutual coherence time between the dual-combs, with a corresponding increase in signal-to-noise ratio (SNR). We refer to this situation as "fully-resolved and accurate" dual-comb spectroscopy. Fully-resolved and accurate dual-comb spectrometers have been demonstrated in the near infrared (NIR) [14, 17, 18]. There have also been important direct comb spectroscopy demonstrations in the mid-infrared (MIR) [9, 10, 13, 19], but not yet in a fully resolved and accurate implementation. Here we demonstrate a fully resolved and accurate MIR dual-comb spectrometer and measure the amplitude and phase spectra of the P, Q and R-branch of methane asymmetric stretch $\nu_3$ band with 45 000 spectral elements spanning 4.5 THz (150 cm$^{-1}$). We report line center determinations across 132 of the stronger $\nu_3$ rovibrational lines. These data contribute to the spectroscopic data on the scientifically and environmentally important methane molecule [23-30] and can be used to calibrate high-resolution FTIR in this spectral region.

As mentioned earlier, while the spectrometer will have a spectral frequency axis with a fractional accuracy set by the comb teeth (which in turn are referenced to a hydrogen maser), this calibration alone does not guarantee a similar accuracy in the line center determination of a molecular transition both because of limited SNR and potential systematics. (This statement is true of not only a dual-comb spectrometer but any direct comb spectrometer.) The most accurate comparison to date for a dual-comb spectrometer was limited to a few MHz by unknown pressure shifts masking possible systematics [14]. The potential for systematic frequency shifts arise from the different use of the comb in direct comb spectroscopy compared to frequency metrology. Optical frequency metrology relies on only the frequency of a single comb tooth, but the dual-comb spectrometer in addition relies on the relative amplitude and phase across multiple comb teeth. Nonlinearities that mix the amplitude and phase of different comb teeth distort the spectral shape, causing effective frequency shifts in molecular line centers and will ultimately limit accuracy. Indeed, we do find photodetection nonlinearities can cause systematic shifts, but we also show that these shifts can be substantially suppressed by acquiring data with both the forward and reverse rf-to-optical mappings shown in Fig. 1. By comparison to saturated absorption spectroscopy of the P(6) and P(7) manifolds [5, 21, 22], we find a systematic uncertainty of ~300 kHz ($10^{-5}$ cm$^{-1}$), or ~1 part per thousand of the Doppler broadened linewidth and a tenfold higher accuracy than previous FTIR data.

## II. MIR DFG DUAL-COMB SPECTROMETER

### A. Experimental Setup

Figure 2 shows the MIR implementation of Fig. 1. Two NIR frequency fiber combs are generated by fs-fiber lasers centered at 1.56 μm where the repetition rates of the sensing comb, $f_{r,S}$, and of the LO comb, $f_{r,L}$, are both about 100 MHz and differ by $\Delta f_r = f_{r,L} - f_{r,S} \approx 1.5$ kHz [17]. The output laser pulses are stretched to picosecond pulsewidths in fiber, amplified to ~0.5 W, combined in a fiber coupler with ~0.5 W of amplified 1.064 μm light from a cw fiber laser, and focused onto periodically poled lithium niobate (PPLN) to generate the MIR frequency combs through difference frequency generation (DFG) [31, 32]. The NIR combs are filtered with an adjustable 7 nm wide bandpass filter to match the phase-matching bandwidth of the 10 mm long PPLN. The generated MIR combs are centered around 3.4 μm with a ~1 THz width and can be scanned over ~4.5 THz (limited by the C-band Er-fiber amplifiers) by adjusting the PPLN temperature and filter position. For frequency measurements from 87 THz to 90 THz, the PPLN poling period was 29.9 μm, and from 90 THz to 92 THz it was 30.2 μm. There are a number of other approaches that can generate MIR combs [8, 19, 33-36] both broader and brighter than the approach taken here, but more development of these sources is needed to establish the very low phase noise across the comb required for fully resolved and accurate, high-SNR dual-comb spectroscopy. A Ge wedge blocks the residual NIR light, leaving 30 μW of MIR light across 7500 teeth, or 4 nW per tooth. The sensing MIR comb passes through a 28 cm long, 200 mTorr methane cell, is heterodyned against the LO MIR comb on an InAs photodiode at a heterodyne efficiency of 30%, and the output is digitized synchronously with $f_{r,L}$.

To achieve high SNR, we require long integration times, which in turn requires high mutual coherence between the sensing and LO combs. Specifically, for the 13-minute acquisition times here, we require an effective relative linewidth below $1/(13\times60$ s$) \approx 1$ mHz, or equivalently in the time domain, sub-radian carrier phase jitter over 13 minutes. We achieve this high mutual coherence at short times (sub-second) through phase-locked loops, as described next, and at longer times (seconds to minutes) through simple software phase correction, as described later. The short-time coherence is maintained by phase-locking both fiber combs tightly to two cavity-stabilized cw transfer lasers as in Ref. [17] so that the mutual linewidths of the two combs is well below 1 Hz. (The 1.064 μm laser is common to both paths and therefore imparts no additional phase noise.)

The mutual coherence allows for long integration times and high SNR, but it does not establish absolute frequency accuracy (as both combs can still move in concert). The frequency accuracy is achieved by recording the absolute frequencies of one cw transfer laser and the 1.064 μm cw laser against a self-referenced frequency comb referenced to a hydrogen maser (H-maser). Knowledge of these frequencies and any rf offsets allows us to determine the anchor frequency, $f_A$ or $f'_A$, as defined in Fig. 1. To extend the frequency accuracy across all the teeth, we record the repetition rates of the combs against a frequency counter, also referenced to the H-maser. The accuracy of the H-maser is $10^{-13}$ [37], which yields a 10 Hz frequency accuracy in the MIR at around 100 THz. Taking into account synchronization between the data acquisition and calibration, as well as any temperature fluctuations, a conservative upper limit is 100 Hz, which is a negligible contribution to our ultimate line center uncertainties.

Unlike conventional FTIR, in a dual-comb spectrometer the frequency resolution and point spacing are decoupled. The frequency resolution (or instrument lineshape) is limited by the *absolute* comb tooth linewidth over the measurement time. In our case the combs are phase locked to two cw reference lasers which are phase locked to an optical cavity (Fig. 2a). All locks exhibit sub-radian phase noise, and the linewidth is dominated by the drift of the optical cavity to 1 kHz to 10 kHz over a typical 13 minute measurement period. This drift could be removed by continuous correction against the self-referenced comb, in which case we would be limited by the ~kHz frequency noise on our 1.064 μm laser (loosely locked to the self-referenced comb) and a similar level of excess frequency noise added to the two cw transfer lasers during fiber-optic transmission to the dual-comb spectrometer. The spectral point spacing is generally much coarser than the resolution and is set by the comb repetition rate ($f_{r,S} \approx 100$ MHz) with finer spacing achievable by step scanning the comb.

### B. Data Acquisition and Coherent Averaging

An example rf spectrum from a continuous data acquisition is given in Fig. 2b. Note that both the amplitude and phase response of the gas are extracted. We measure a well resolved rf comb with sub-Hertz linewidth, despite the the addition of fiber optics, optical filters, DFG optics, and Er-doped fiber amplifiers compared to Ref. [14]. These data were acquired without any software phase correction, relying instead only on the mutual coherence established between the two combs through the phase locked loops. Continuous, uninterrupted data acquisition is limited by the size of the digitizer memory to about 5 seconds.

High SNR requires averaging times longer than this 5 seconds. To obtain long averaging times with zero dead time and without cumbersome file sizes, we implement coherent, real-time co-adding of sequential interferograms [17, 18]. This co-adding requires the phase of successive interferograms to be identical, to within any noise, which in turn requires two conditions. First, the combs should be phase locked to coincide at an anchor frequency $f_A$ (or $f'_A$) (Fig. 1b) and second the repetition rates of the LO and sensing comb, $f_{r,L}$ and $f_{r,S}$ must satisfy $(1+M)f_{r,L} = Mf_{r,S}$, where $M$ is an integer. The phaselocks have been improved relative to Ref. [18] and can be tuned so that these conditions can be met as the MIR comb is scanned across 4.5 THz with $M = 2^{16}$. Simultaneously $f_A$ (or $f'_A$) is adjusted to remain ~1.2 THz below (or above) the filtered comb carrier frequency so that the rf comb remains centered at $(\Delta f_r/f_{r,L}) \times (1.2 \text{ THz}) = 18$ MHz. For this case, each interferogram is exactly $2^{16}$ points long and exactly repeats at $\Delta f_r \approx 1.5$ kHz. 100 sequential inteferograms are added in real time over 0.07 s in a field programmable gate array (FPGA), limited by slow carrier phase wander. For longer times, these summed interferograms are phase-corrected by removing only the linear component of the spectral phase (as opposed to the more extensive phase correction used in conventional FTIR [38]) to generate a final high-SNR interferogram. This interferogram is Fourier transformed to yield the complex rf spectrum with points spaced by $f_{r,S}$, at a resolution of 1 kHz to 10 kHz, and with a frequency accuracy of better than 100 Hz. This process of summing in time and then Fourier transforming the data provides exactly the same results as Fourier transforming the entire time data record and then selecting the amplitude and phase at each distinct rf tooth (e.g. as in Fig. 2c), except that it is dramatically faster and less memory intensive.

## III. RESULTS

### A. Complex (Phase and Amplitude) Spectra and Line Centers

In terms of the molecular absorption coefficient, $\alpha(f)$, and phase delay $\varphi(f)$, the measured complex spectrum is $S(f) = S_0(f) \exp[-\alpha(f)L/2 - i\varphi(f)]$, where $L$ is the cell length and $S_0(f)$ is the complex heterodyne signal between the combs in the absence of the sample. Figure 3 shows complex methane spectra across 4.5 THz, each acquired over 13 minutes as the temperature of the PPLN was stepped to access different spectral regions.

The statistical spectral SNR for the amplitude or phase is limited by the InAs detector's ~300 µA dark current to ~125/√s, or ~3500 for a full 13 minute acquisition near the center of the spectrum, with an average value across the full width at half maximum (FWHM) of around 100/√s. This SNR for the magnitude and phase of the complex spectrum corresponds as well to the statistical SNR for the measured molecular absorbance divided by two, $\alpha(f)L/2$, and the molecular-induced phase delay, $\varphi(f)$, in radians. The product of SNR and number of spectral elements is $10^6/\sqrt{s}$, nearly the same dynamic-range limited value as for previous NIR dual-comb systems [14, 18, 39]. Although higher optical power would in principle improve the SNR, it would also increase nonlinear effects and therefore systematic shifts, as discussed in the next section. Variations across the baseline on the frequency scale of the molecular features can further degrade the SNR beyond this statistical limit.

To isolate the molecular response, we compute $-\ln[S(f)] = \alpha(f)L/2 + i\varphi(f) - \ln[S_0(f)]$. The last term includes the more slowly varying baseline wander. The first two terms are the absorbance and phase responses from all rovibrational lines, each of which is proportional to the complex error function $w(x) = \exp(-x^2) + i2\pi^{-1/2}F(x)$ for pure Doppler broadening with the scaled frequency $x = (f - f_k)/\sigma_D$, line center $f_k$ and Doppler width $\sigma_D$. The real (absorbance) part is the usual Gaussian profile and the imaginary (phase) part is the corresponding Dawson integral, $F(x)$. We fit both the real (absorbance) and imaginary (phase) parts of $S(f)$ separately over a 2 GHz to 20 GHz window around each line, or lines, using a 3rd order polynomial for the local baseline variations. Examples of fits for the P(6) and P(7) transitions are given in Fig. 3b along with the fit residuals. There is good agreement between the absorbance and phase fits in terms of fitted line centers.

One concern is the relatively sparse sampling of 100 MHz and its impact on the line center determination. In principle, since the Gaussian or Dawson fitting function has only three free parameters, as few as three measured spectral points with high SNR are sufficient for a line center determination. For the Gaussian fit, the statistical uncertainty on the fitted line center will be $\sigma_{fk} \sim \Delta\nu_D / \left(SNR_k \sqrt{N_{pnts}}\right)$ [38], where $SNR_k$ is the SNR of the $k_{th}$ peak, $\Delta\nu_D = 2\sqrt{2\ln(2)}\sigma_D$ is its Doppler FWHM and $N_{pnts} = \Delta\nu_D / f_{r,S}$ is the number of points across the peak. For a $\Delta\nu_D$ of 300 MHz, a point spacing of $f_{r,S} = 100$ MHz, and an SNR of 3500, $\sigma_{fk} \sim 50$ kHz. The fit uncertainties to the phase profiles depend similarly on SNR and width. This estimate is consistent with the numerical uncertainties returned by the fit. To establish that the 100 MHz sampling does not cause any systematic shifts, we acquired four "interleaved" spectra by step-scanning the 1.064 μm laser frequency in 25 MHz steps for the P(7) and P(6) transitions, shown in Fig. 4. The fit results for the manifold of lines are statistically identical across the four data sets verifying that the 100 MHz sampling, while sparse, is sufficient where the fit function is known and was used for the remainder of the lines. In addition these data indicate the potential of this method for line shape studies.

While the SNR is sufficient to observe lines (at 3σ) with line strengths above $3.7 \times 10^{-23}$ cm$^{-1}$/(molecule cm$^{-2}$) for the 28 cm long cell, we restrict the fits for the line center determinations to 132 single lines with strength $>3.7 \times 10^{-22}$ cm$^{-1}$/(molecule cm$^{-2}$) over the P, Q and R(0)-R(2) branch. The line centers given in Table I are calculated from the average over the phase and absorbance fits (and over the two different rf-to-optical mappings as described in the next section.) The averaged fit uncertainties depend on line strength and existence of any adjacent lines; they are below 30 kHz for strong lines and below 300 kHz ($10^{-5}$ cm$^{-1}$) for 106 of the 132 fitted lines. The relative line strengths match Ref. [28] to within 1% across the P-branch.

Methane is a well-studied molecule and there has been previous high-accuracy spectroscopy of the ν3 band. Specifically, the P(6) and P(7) lines have been well characterized with a comb-based saturated absorption spectrometer [5, 21, 22]. In Fig. 5, we plot the difference between our measurements and the saturated absorption data, which has an uncertainty of ~10 kHz. The maximum difference across all the lines is 275 kHz, or $10^{-3}$ of the Doppler broadened linewidth. Based on this comparison and the discussion in the next section, we assign an overall systematic uncertainty of 300 kHz to the line centers of Table I. In addition, we can compare the entire data set to previous values in the literature taken with high-resolution FTIR [28, 42], as shown in Fig. 6. For the most part, the FTIR data is within its specified accuracy of ~ 3 MHz.

**B. Evaluation of Systematic Shifts**

While the frequencies of the sample points are accurate and the fit uncertainties can be below 30 kHz, systematics can shift the measured line centers and must be considered. The effect of the baseline wander is included in the fitted uncertainty. (Since the signals at all optical frequencies are measured simultaneously, there is no time-dependent baseline wander as can arise in swept laser measurements.) The heterodyne signal occurs at well-defined rf frequencies spanning 15 MHz to 30 MHz, well away from $1/f$ noise, with an rf bandwidth much narrower than the 50 MHz Nyquist frequency (Fig. 2b) so there are no extraneous backgrounds or aliasing effects. As discussed above, the frequency axis is very well defined and contributes negligible error.

Light shifts will be minimal since the intense NIR light is removed prior to the gas cell and the MIR light has a peak pulse intensity <1 W and broad spectrum. Self-pressure shifts can occur at our 200 mTorr pressure. They are known and can be corrected for the Q-lines [29] but have not been measured across the P or R branches. The corresponding shifts from 200 mTorr of $N_2$ would be around −50 kHz [40]. Assuming the self-pressure shifts are double the $N_2$-induced shifts, as is the case for the Q-branch [29], yields a potential systematic shift of around −100 kHz.

As mentioned earlier, nonlinearities that mix the amplitude or phase of the sensing comb teeth *after* transmission through the methane can systematically distort the molecular signature. This systematic does not appear in frequency metrology as the comb tooth frequencies are unaffected; it appears here because the comb tooth amplitude and phase, as well as its frequency, carry the signal. Considering the low peak powers and minimal optical path overlap, optical nonlinearities are negligible. The same, however, is not true in the rf-domain and particularly in the photodetection. We find a line-center shift of around ±1 MHz for the phase and amplitude that depends on whether $f_A$ is positioned 1.2 THz below or above the center of the MIR comb (Fig. 1b), corresponding to a forward or reverse mapping of the optical to the rf domain (and sign reversal of any nonlinear-induced shifts). Any real model of this systematic is quite complicated as it depends on the response of the photodetection to both incident pulse trains. This basic effect has been modeled in the context of low phase noise microwave phase generation by photodetection of a comb with some success [41], but the dual-comb system would require an even more sophisticated model of the detector response. Nevertheless, we can suppress this systematic by acquiring data in each mapping at roughly the same power levels (to within 15%) and calculating the average. It is this average that is reported in Table I.

As discussed in the previous Section, the ultimate evaluation of systematic shifts is based on the comparison of Fig. 5. Based on these data and since the remainder of the data were acquired under almost identical conditions, we assign 300 kHz as the overall systematic uncertainty.

## IV. CONCLUSION

These data represent the first accurate line center measurements made with a coherent dual-comb spectrometer and demonstrate some of its features relative to conventional spectroscopy. Compared to conventional high-resolution FTIR, the dual-comb spectrometer measures both phase and absorbance and provides line centers with about ten times better accuracy. Compared to Doppler-free saturated absorption spectroscopy, the dual-comb spectrometer has an order of magnitude lower accuracy, but it completely avoids the demanding requirement of a high power laser or build-up cavity to saturate the transition while also covering a broader spectral range. Additionally, the accuracy of this spectrometer will be retained when measuring collisionally broadened spectra. The SNR per root second is low compared to cw laser spectroscopy, but coherent averaging yields high SNRs over many spectral elements (equal to 45 000 here). This work focused on center frequency determination of Doppler-broadened lines at the level of $3\times10^{-9}$ fractional accuracy, but the ability to measure both phase and amplitude as well as the quality of the fits indicate the potential for line shape or line mixing studies, which remain an experimentally challenging problem with important implications to greenhouse gas monitoring.

This work was funded by the National Institute of Standards and Technology (NIST). F.R.G. received support from the Swiss National Science Foundation (SNF) under grant no. PBNEP2-127797. The authors acknowledge helpful discussions with F. Adler, L. Brown and L. Nugent-Glandorf.

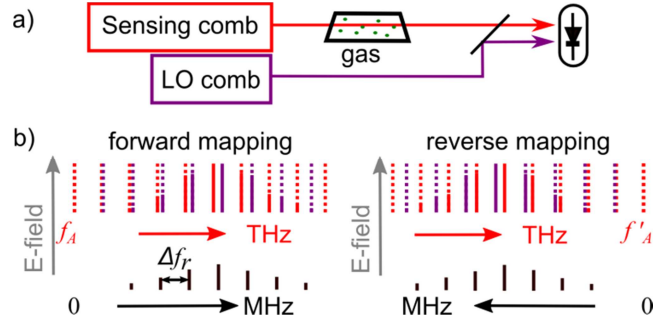

FIG. 1. (a) Dual-comb spectrometer for amplitude *and phase* measurements of a gas sample by use of a sensing comb that passes through the gas and a local oscillator (LO) comb with a repetition rate that differs by $\Delta f_r$. (b) One-to-one mapping of the sensing comb teeth (red lines) to the rf domain achieved by heterodyning against the LO comb (purple lines) over the comb spectra (solid lines). The combs are phase-locked such that they overlap at the anchor tooth frequency, $f_A$, just below the optical spectrum for forward mapping (left side) or at $f'_A$ just above the spectrum for reverse mapping (right side). Data is acquired with both forward and reverse mapping to reduce systematics associated with photodetection nonlinearities. Note that only one comb transverses the sample in this configuration; if both do, the absorbance cannot be ascribed to a single tooth at a single frequency and the phase delay from the gas is not measured.

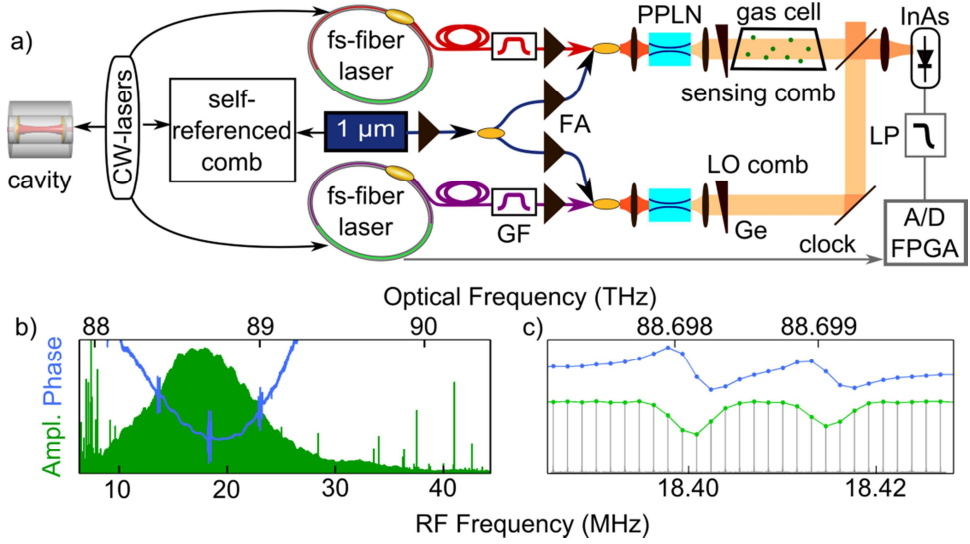

FIG. 2. (a) Coherent MIR dual-comb spectrometer for measurement of methane around 3.4 µm. GF: optical grating filter, FA: fiber amplifier, Ge: Germanium wedge, LP: 50 MHz low pass filter, A/D: Analog-to-Digital converter, CW-lasers: 1.535 µm and 1.560 µm cw fiber lasers, 1 µm: 1.064 µm cw fiber laser. (b) Measured rf amplitude (green) and phase (blue) versus rf frequency (bottom scale) and calculated optical frequency (top scale) over a ~50 MHz rf window. The phase data shows the phase profile from the P(5) through P(7) transitions superimposed on the differential chirp (parabola) between the combs. A few spurious rf lines near 10 MHz are fully rejected since they do not fall on the rf comb grid (i.e. at harmonics of $\Delta f_r$). (c) An expanded 40 kHz rf span that clearly shows the discrete comb structure with a tooth separation of $\Delta f_r$ = 1.5 kHz in the rf or $f_{r,S}$= 100 MHz in the optical, as in Fig. 1b. The observed features are then P(6) $F_2^{(1)}$ and $E^{(1)}$ lines. As in (b) the phase is only sensible, and plotted, at the comb teeth positions. These data correspond to 1.34 s of continuous data acquisition and each rf tooth has a time-bandwidth limited linewidth of 0.75 Hz, without any software phase correction. For times much beyond a few seconds, the linewidths are broadened but a simple linear software phase correction can maintain the time-bandwidth limited linewidth across the rf comb, for example achieving 0.19 Hz linewidths over 5.3 s.

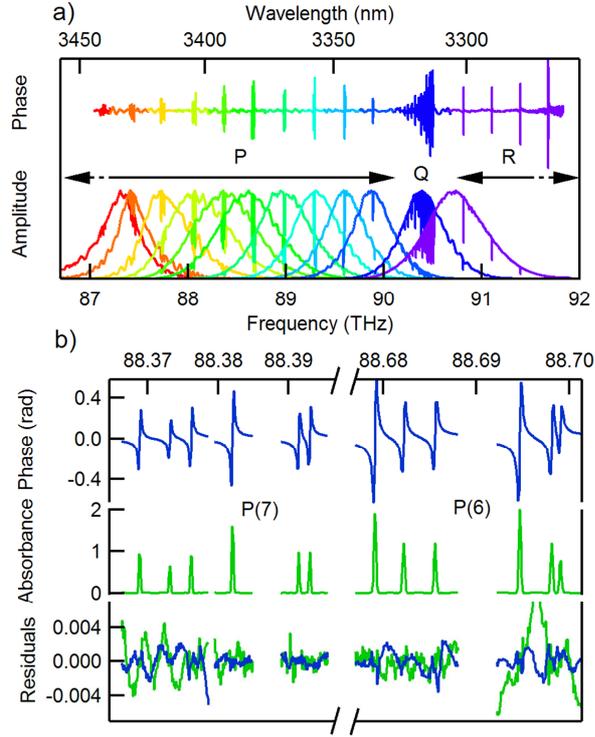

FIG. 3. (a) Normalized amplitude and phase spectra (linear scale) across the P, Q, and partial R-branch. The baseline for the phase spectra was removed by a polynomial fit. The peak SNR is ~3500 in amplitude or phase. (b) The corresponding absorbance, $\alpha(f)L$ (green), and phase, $\phi(f)$ (blue) after a 3$^{rd}$ order polynomial baseline fit, for the P(7) $F_1^{(2)}$, $E^{(1)}$, $F_2^{(2)}$, $A_2^{(1)}$, $F_2^{(1)}$, $F_1^{(1)}$ symmetry lines and the P(6) $A_1^{(1)}$, $F_1^{(1)}$, $F_2^{(2)}$, $A_2^{(1)}$, $F_2^{(1)}$, $E^{(1)}$ symmetry lines. The fit residuals for the absorbance (green) and phase (blue) have standard deviations of $1.3 \times 10^{-3}$ and $1.1 \times 10^{-3}$ respectively, including both statistical and residual baseline noise.

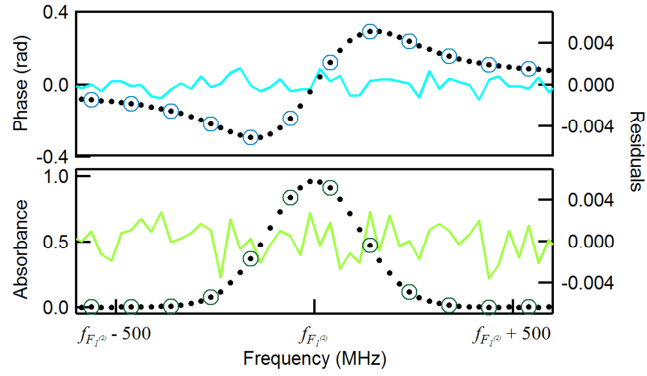

FIG. 4: Example interleaved measurement of the $F_1^{(2)}$ line of P(7), comprising four distinct sets of absorbance and phase spectra. For each set, the 1.064 μm laser was stepped by 25 MHz (solid circles), so that the combined data have one quarter the normal spectral spacing of $f_{r,S}=$ 100 MHz (open circles). The fits demonstrate good agreement (residuals, solid line) with the expected Dawson and Gaussian functions.

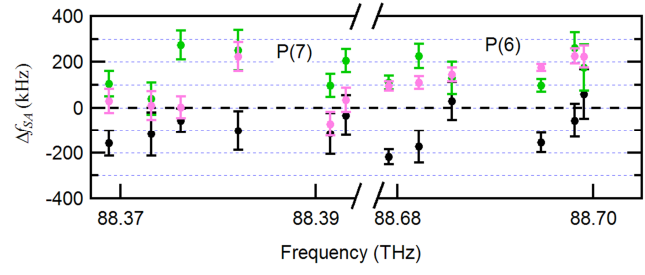

FIG. 5. Difference, $\Delta f_{SA}$, between our measured P(6) and P(7) line centers and saturated-absorption data [5, 21, 22] for three data sets taken several weeks apart (green, pink, and black circles). The error bars reflect the fit uncertainties. The mean offset is +58 kHz and the maximum deviation is +275 kHz.

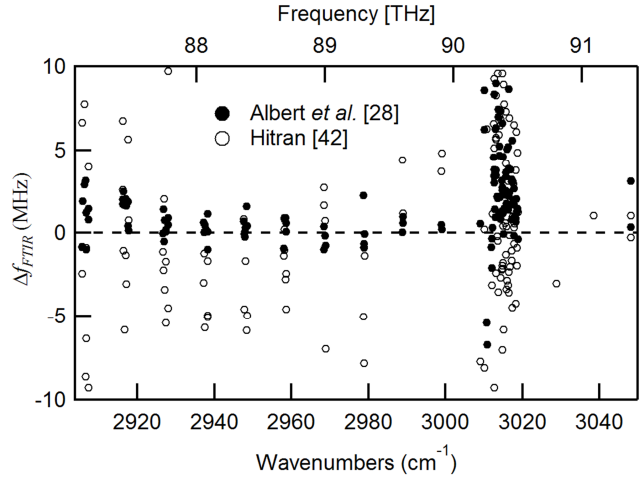

FIG. 6. Difference, $\Delta f_{FTIR}$, between our measurement and FTIR measurements from Refs. [28] and [42] for the P(1) to P(11) lines (87 THz to 90.2 THz), the stronger Q(n) lines (90.2 THz to 90.5 THz) and the R(0) to R(2) lines (90.8 THz to 91.5 THz). $\Delta f_{FTIR}$ is generally below the 3 MHz quoted uncertainty of the FTIR data. (Four lines of [28] and 11 lines of [42] have >10 MHz difference and are off scale.)

Table I: Measured line center frequencies of the methane $\nu_3$ band calculated from the average of the phase and amplitude fits measured under both forward and reverse rf-to-optical mappings. The data for P(11) to P(2) is the average of two separate data sets, while the remainder is from a single data set. Comp.: lower level component $JC^{(\alpha)}$; $J$: rotational quantum number, $C$: $T_d$ point group symmetry, $\alpha$: multiplicity index. $f_k$ (MHz): measured line center frequency. $\sigma$ (MHz): statistical uncertainty on $f_k$ estimated from fit residual, divided by two to account for the averaging across phase and amplitude fits and across both mappings. $\Delta_A$ (MHz): difference between our $f_k$ and FTIR data from Albert et al. [28]. $\Delta_H$ (MHz): difference between our $f_k$ and data from Hitran [42]. In addition to the fit uncertainties, $\sigma$, we assign a systematic uncertainty of 0.30 MHz as discussed in the text.

| Comp. | $f_k$ (MHz) | $\sigma$ | $\Delta_A$ | $\Delta_H$ | Comp. | $f_k$ (MHz) | $\sigma$ | $\Delta_A$ | $\Delta_H$ |
|---|---|---|---|---|---|---|---|---|---|
| | **P(11)** | | | | 7 $F_2^{(1)}$ | 88391450.56 | 0.09 | 1.59 | -5.85 |
| 11 $F_1^{(3)}$ | 87108705.86 | 0.22 | -0.80 | -2.45 | 7 $F_1^{(1)}$ | 88393030.08 | 0.09 | 0.42 | -4.94 |
| 11 $E^{(2)}$ | 87110620.58 | 0.37 | -0.85 | 6.64 | | **P(6)** | | | |
| 11 $F_2^{(3)}$ | 87114104.29 | 0.24 | 1.91 | 15.40 | 6 $A_1^{(1)}$ | 88679126.58 | 0.03 | -0.90 | -1.35 |
| 11 $F_1^{(2)}$ | 87128156.56 | 0.27 | 2.91 | 7.76 | 6 $F_1^{(1)}$ | 88682207.80 | 0.07 | 0.91 | -1.01 |
| 11 $E^{(1)}$ | 87137341.07 | 0.37 | 3.15 | -8.63 | 6 $F_2^{(2)}$ | 88685592.40 | 0.08 | 0.86 | 0.23 |
| 11 $F_2^{(2)}$ | 87139103.66 | 0.29 | -1.00 | -0.88 | 6 $A_2^{(1)}$ | 88694688.75 | 0.04 | 0.90 | -2.78 |
| 11 $A_2^{(1)}$ | 87141721.91 | 0.15 | 1.24 | -6.35 | 6 $F_1^{(1)}$ | 88698120.57 | 0.07 | 0.10 | -4.58 |
| 11 $F_2^{(1)}$ | 87159299.14 | 0.25 | 1.46 | -9.27 | 6 $E^{(1)}$ | 88699071.40 | 0.11 | 0.59 | -2.44 |
| 11 $F_1^{(1)}$ | 87159743.49 | 0.25 | 0.83 | 3.98 | | **P(5)** | | | |
| | **P(10)** | | | | 5 $F_1^{(2)}$ | 88990490.75 | 0.06 | 0.39 | 1.68 |
| 10 $F_2^{(3)}$ | 87425525.14 | 0.10 | 2.01 | 2.61 | 5 $E^{(1)}$ | 88992600.75 | 0.08 | -1.00 | 2.75 |
| 10 $E^{(2)}$ | 87428529.63 | 0.12 | 1.74 | 6.75 | 5 $F_2^{(1)}$ | 89000472.61 | 0.04 | -0.16 | 0.74 |
| 10 $F_1^{(2)}$ | 87431362.24 | 0.11 | 2.51 | -1.06 | 5 $F_1^{(1)}$ | 89004040.23 | 0.04 | -0.73 | -6.96 |
| 10 $A_1^{(1)}$ | 87442078.68 | 0.06 | 1.67 | -5.82 | | **P(4)** | | | |
| 10 $F_1^{(1)}$ | 87448447.19 | 0.10 | 2.05 | -1.34 | 4 $F_2^{(1)}$ | 89297695.06 | 0.04 | 2.27 | -5.02 |
| 10 $F_2^{(2)}$ | 87451442.36 | 0.08 | 1.63 | -3.04 | 4 $E^{(1)}$ | 89303619.87 | 0.08 | -0.63 | -0.87 |
| 10 $A_2^{(1)}$ | 87468319.42 | 0.10 | 1.87 | 5.65 | 4 $F_1^{(1)}$ | 89305771.40 | 0.05 | -0.83 | -7.82 |
| 10 $F_2^{(1)}$ | 87469023.39 | 0.37 | 0.44 | -11.37 | 4 $A_1^{(1)}$ | 89308512.19 | 0.03 | -0.05 | -1.37 |
| 10 $E^{(1)}$ | 87469331.90 | 0.56 | 0.16 | 0.79 | | **P(3)** | | | |
| | **P(9)** | | | | 3 $A_2^{(1)}$ | 89601828.58 | 0.04 | 0.06 | 4.41 |
| 9 $A_2^{(1)}$ | 87740265.55 | 0.08 | -0.03 | -1.14 | 3 $F_2^{(1)}$ | 89605943.43 [a] | 0.10 | 0.98 | 1.19 |
| 9 $F_2^{(2)}$ | 87742746.60 | 0.08 | 1.44 | -2.22 | 3 $F_1^{(1)}$ | 89608969.54 | 0.04 | 0.59 | 10.81 |
| 9 $F_1^{(3)}$ | 87745807.05 | 0.24 | -0.49 | 2.06 | | **P(2)** | | | |
| 9 $A_1^{(1)}$ | 87751536.56 | 0.07 | 0.77 | -3.40 | 2 $F_2^{(1)}$ | 89907579.08 | 0.08 | 0.50 | 3.71 |
| 9 $F_1^{(2)}$ | 87760424.08 | 0.16 | 0.19 | -5.39 | 2 $E^{(1)}$ | 89909569.43 | 0.12 | 0.23 | 4.81 |
| 9 $E^{(1)}$ | 87762118.53 | 0.24 | 0.09 | -1.71 | | **P(1)** | | | |
| 9 $F_2^{(1)}$ | 87777197.04 | 0.09 | 0.93 | -4.50 | 1 $F_1^{(1)}$ | 90207892.04 [b] | 0.45 | 0.56 | -7.71 |
| 9 $F_1^{(1)}$ | 87778139.32 | 0.09 | 0.51 | 9.71 | | **Q** | | | |
| | **P(8)** | | | | 14 $A_1^{(1)}$ | 90257623.06 | 0.58 | -5.42 | 6.27 |
| 8 $F_2^{(2)}$ | 88056079.37 | 0.08 | 0.63 | -3.00 | 13 $F_1^{(2)}$ | 90264292.48 [c,d] | 8.52 | -25.07 | 16.03 |
| 8 $E^{(2)}$ | 88058283.86 | 0.25 | 0.06 | -1.23 | 13 $A_1^{(1)}$ | 90264855.74 | 0.38 | -6.70 | 14.73 |
| 8 $F_1^{(2)}$ | 88063885.76 | 0.07 | 0.49 | -5.69 | 13 $F_2^{(2)}$ | 90295381.21 | 0.54 | -0.86 | 20.13 |
| 8 $F_2^{(1)}$ | 88072045.73 | 0.07 | 0.14 | 0.23 | 12 $F_2^{(1)}$ | 90298367.52 | 0.51 | -2.10 | 10.46 |
| 8 $E^{(1)}$ | 88084799.29 | 0.15 | 0.10 | -4.96 | 12 $F_1^{(2)}$ | 90299371.08 | 0.52 | -0.32 | -3.11 |
| 8 $F_1^{(1)}$ | 88085481.90 | 0.10 | 1.17 | -5.07 | 13 $A_2^{(1)}$ | 90304361.13 | 0.29 | 0.34 | 1.14 |
| 8 $A_1^{(1)}$ | 88086573.91 | 0.06 | -0.97 | -1.69 | 13 $F_2^{(3)}$ | 90312583.87 | 2.79 | 4.56 | 6.57 |
| | **P(7)** | | | | 13 $F_1^{(4)}$ | 90318786.02 | 0.86 | 3.41 | 5.12 |
| 7 $F_1^{(2)}$ | 88368863.33 | 0.06 | 0.72 | 0.84 | 12 $E^{(2)}$ | 90323056.40 | 0.78 | 1.45 | 20.30 |
| 7 $E^{(1)}$ | 88373148.98 | 0.10 | -0.07 | -4.59 | 12 $F_2^{(2)}$ | 90324951.19 | 0.53 | 0.95 | 6.26 |
| 7 $F_2^{(2)}$ | 88376181.71 | 0.05 | -0.21 | 0.09 | 11 $A_2^{(1)}$ | 90329810.65 | 0.13 | 3.48 | -2.40 |
| 7 $A_2^{(1)}$ | 88382052.81 | 0.09 | 0.32 | -1.69 | 11 $F_2^{(2)}$ | 90331277.00 | 0.23 | 3.78 | 5.76 |

| Comp. | $f_k$(MHz) | $\sigma$ | $\Delta_A$ | $\Delta_H$ | Comp. | $f_k$(MHz) | $\sigma$ | $\Delta_A$ | $\Delta_H$ |
|---|---|---|---|---|---|---|---|---|---|
| Q continued | | | | | 8 $F_2^{(2)}$ | 90428469.93 | 0.14 | 5.19 | 3.12 |
| 11 $E^{(1)}$ | 90332131.19 | 0.58 | 9.02 | 8.27 | 6 $E^{(1)}$ | 90431140.15 | 0.21 | 2.22 | -3.59 |
| 12 $A_2^{(1)}$ | 90332479.75 [e] | 0.38 | 6.34 | 5.65 | 6 $F_2^{(1)}$ | 90432068.53 | 0.78 | 8.65 | -2.33 |
| 12 $F_2^{(3)}$ | 90339694.78 | 0.28 | 2.19 | -0.18 | 7 $A_2^{(1)}$ | 90432325.38 | 0.57 | 3.88 | -10.84 |
| 12 $F_1^{(3)}$ | 90342803.94 | 0.49 | 2.08 | 9.60 | 6 $A_2^{(1)}$ | 90434355.20 | 0.09 | 2.42 | -0.13 |
| 12 $A_1^{(2)}$ | 90345235.40 | 0.24 | 2.10 | -0.48 | 7 $F_2^{(2)}$ | 90436607.56 | 0.14 | 1.72 | 6.93 |
| 9 $F_1^{(1)}$ | 90348788.81 | 0.12 | 6.99 | -3.53 | 7 $E^{(1)}$ | 90439384.24 | 0.10 | 1.72 | -2.03 |
| 9 $F_2^{(1)}$ | 90349166.95 | 0.12 | 7.45 | -15.42 | 7 $F_1^{(2)}$ | 90445574.72 | 0.05 | 3.82 | -11.05 |
| 11 $F_1^{(2)}$ | 90351646.49 | 0.22 | 1.23 | 5.94 | 5 $F_1^{(1)}$ | 90452276.50 | 0.12 | 0.81 | -1.05 |
| 11 $F_2^{(3)}$ | 90359262.32 | 0.24 | 4.67 | 0.92 | 6 $F_2^{(2)}$ | 90454220.44 | 0.14 | 3.11 | -1.63 |
| 10 $F_2^{(2)}$ | 90360309.77 | 0.15 | 5.24 | 7.04 | 5 $F_2^{(1)}$ | 90455379.92 | 0.12 | 3.24 | 1.35 |
| 10 $F_1^{(1)}$ | 90362596.90 | 0.19 | 2.15 | 6.49 | 6 $F_1^{(1)}$ | 90457753.43 | 0.05 | 5.57 | -4.47 |
| 11 $E^{(2)}$ | 90365736.74 | 0.36 | 7.35 | -2.66 | 6 $A_1^{(1)}$ | 90461384.72 | 0.04 | 3.08 | 1.01 |
| 11 $F_1^{(3)}$ | 90367765.34 | 0.24 | 0.85 | 6.87 | 4 $A_1^{(1)}$ | 90468721.00 | 0.04 | 1.99 | 2.95 |
| 10 $A_1^{(1)}$ | 90376685.24 | 0.10 | 3.15 | -2.19 | 5 $E^{(1)}$ | 90470260.15 | 0.09 | 1.71 | 6.50 |
| 8 $A_1^{(1)}$ | 90378783.67 | 0.05 | 2.76 | -2.13 | 4 $F_1^{(1)}$ | 90471813.60 | 0.14 | -0.17 | -0.65 |
| 8 $F_1^{(1)}$ | 90379477.61 | 0.12 | 2.59 | -1.97 | 5 $F_1^{(2)}$ | 90472131.82 | 0.15 | 2.67 | 0.33 |
| 8 $E^{(1)}$ | 90379843.29 | 0.18 | 6.60 | -0.70 | 4 $E^{(1)}$ | 90473934.30 | 0.09 | 1.00 | 0.58 |
| 10 $F_1^{(2)}$ | 90382579.48 | 0.32 | 1.34 | 9.58 | 4 $F_2^{(1)}$ | 90483521.64 | 0.05 | 0.67 | -4.24 |
| 10 $E^{(2)}$ | 90385664.90 | 0.55 | 4.62 | 0.42 | 3 $F_1^{(1)}$ | 90484622.99 | 0.05 | 0.89 | 3.77 |
| 9 $E^{(1)}$ | 90387469.83 | 0.37 | 1.18 | -1.76 | 3 $F_2^{(1)}$ | 90488114.34 | 0.04 | 2.06 | 6.10 |
| 9 $F_1^{(2)}$ | 90388941.18 | 0.25 | -0.06 | -5.81 | 3 $A_2^{(1)}$ | 90493215.50 | 0.03 | 1.34 | -0.88 |
| 10 $F_2^{(3)}$ | 90390172.00 | 0.36 | 1.37 | 8.93 | 2 $E^{(1)}$ | 90495092.15 | 0.08 | 1.30 | -1.94 |
| 9 $A_1^{(1)}$ | 90393517.84 | 0.06 | 1.46 | 7.76 | 2 $F_2^{(1)}$ | 90496856.61 | 0.05 | 1.48 | 4.87 |
| 9 $F_1^{(3)}$ | 90405139.22 | 0.27 | 2.29 | 0.61 | 1 $F_1^{(1)}$ | 90502080.75 | 0.05 | -0.36 | 1.26 |
| 7 $F_1^{(1)}$ | 90406665.10 | 0.14 | 2.53 | -1.28 | R(0) | | | | |
| 7 $F_2^{(1)}$ | 90407889.51 | 0.15 | 3.64 | 7.29 | 0 $A_1^{(1)}$ | 90799708.38 | 0.07 | -19.27 | -3.02 |
| 9 $F_2^{(2)}$ | 90408599.89 | 0.27 | 3.42 | 4.20 | R(1) | | | | |
| 9 $A_2^{(1)}$ | 90411565.52 | 0.21 | 13.67 | -3.36 | 1 $F_1^{(1)}$ | 91091893.24 | 0.10 | -21.14 | 1.05 |
| 8 $F_2^{(1)}$ | 90411883.73 | 0.25 | 1.78 | 0.40 | R(2) | | | | |
| 8 $F_1^{(2)}$ | 90417079.29 | 0.09 | 5.05 | -2.87 | 2 $F_2^{(1)}$ | 91381337.88 | 0.27 | 0.36 | -0.25 |
| 8 $E^{(2)}$ | 90426450.83 | 0.21 | 1.39 | -3.14 | 2 $E^{(1)}$ | 91381810.83 | 0.41 | 3.14 | 1.04 |

[a–e] For the fits to these lines, there were additional weaker overlapping lines (listed in [28]) that do not appear in the table but that were included in the fit to avoid pulling of the line centers. The weaker lines included are [a] 8 $F_1^{(1)}$, [b] 10 $A_1^{(1)}$, [c] 15 $A_2^{(2)}$, [d] 13 $E^{(1)}$ and [e] 10 $A_1^{(1)}$.